\documentclass[preprint,12pt]{aastex}
\shorttitle{SMC-Type Interstellar Dust in the Milky Way}

\begin{document}

\title{SMC-Type Interstellar Dust in the Milky Way}
\author{Lynne A.\ Valencic\altaffilmark{1},Geoffrey C.\ 
Clayton\altaffilmark{1}, Karl Gordon\altaffilmark{2}, \&
Tracy L.\ Smith\altaffilmark{3}}
\altaffiltext{1}{Department of Physics \& Astronomy, Louisiana State
University, Baton Rouge, LA 70803; valencic@phys.lsu.edu; gclayton@fenway.
phys.lsu.edu}
\altaffiltext{2}{Steward Observatory, University of Arizona, Tucson, AZ
85721; kgordon@as.arizona.edu}
\altaffiltext{3}{Space Science Institute Columbus, Smith Laboratory, Department of Physics, Ohio State University, Columbus, OH 43210; tsmith@campbell.mps.ohio-state.edu}

\newpage

\begin{abstract}
It is well known that the sightline toward HD 204827 in the
cluster Trumpler 37 shows 
an UV extinction curve that does not follow CCM.  However, when a 
dust component, foreground to the cluster, is removed, the residual 
extinction curve is identical to that found in the SMC within the
uncertainties. The curve is very steep and has little or no 2175 \AA\
bump. The position of HD 204827 in the sky is projected onto the edge of the
Cepheus IRAS Bubble. In addition, HD 204827 has an IRAS bowshock indicating 
that it may be embedded in dust swept up by the supernova that created the
IRAS Bubble. Shocks due to the supernova may have led to substantial
processing of this dust. 
The HD 204827 cloud is dense and rich in carbon molecules.
The 3.4 $\mu$m feature indicating a C-H grain 
mantle is present in the dust toward HD 204827. 
The environment of the HD 204827 cloud dust may be similar to the dust
associated with HD 62542 which lies on the edge of a stellar wind bubble
and is also dense and rich in molecules.
This sightline may be a Rosetta Stone if its environment can be related to
those in the SMC having similar dust.
\end{abstract}


\section{Introduction}

The Cardelli, Clayton, \& Mathis (1989, hereafter CCM) average Milky Way 
extinction relation, A$_{\lambda}$/A$_{V}$, is applicable to a wide range of 
interstellar dust environments, 
including lines of sight through diffuse dust and dark cloud dust, as 
well as dust associated with star formation. 
However, the CCM relation does not usually apply beyond the Milky Way,   
even in other Local Group galaxies such as the Magellanic Clouds 
and M31 (e.g., Clayton \& Martin 1985; Fitzpatrick 1985, 1986; 
Clayton et al. 1996; Bianchi et al. 1996; Gordon \& Clayton 1998; 
Misselt, Clayton, \& Gordon 1999). There is some evidence that 
it may apply along some sightlines in the Large Magellanic Cloud (Gordon et al. 2003).
It is important to understand why dust in these other galaxies is different 
since many extragalactic environments seem to contain interstellar dust 
that is better represented by dust in the SMC than the Milky Way 
(e.g., Gordon, Calzetti, \& Witt 1997; 
Pitman, Clayton, \& Gordon 2000). 
Real deviations from CCM are seen for a few sightlines in the Galaxy 
(Cardelli \& Clayton 1991; Mathis \& Cardelli 1992). Sightlines such as those 
toward 
HD 62542, HD 204827, and HD 210121 show weak bumps and anomalously strong 
far-UV extinction for their measured values of R$_V$. Other steep far-UV, 
weak bump dust was found along some low density, low extinction sightlines
(Clayton, Gordon, \& Wolff 2000). The extinction along these sightlines resembles 
those seen in the LMC but
none approach the extreme properties of the SMC sightlines. Nevertheless,
studying these anomalous Galactic sightlines may be a key to relating the
differing extinction characteristics to  various dust environments seen
in the Milky Way and other galaxies. 

The ultraviolet extinction properties of dust toward 18 stars in Trumpler 37, 
including HD 204827, 
were studied more than 15 years ago (Clayton \& Fitzpatrick 1987, hereafter CF). At that time, 
the UV extinction,
in this region of the sky, was referred to as
anomalous as it was generally steeper than the average Galactic extinction 
curve.  
This extinction in the Trumpler 37 region is no longer
considered anomalous. The extinction curves, with one exception,  
fit the CCM relation with R$_V$ values less
than the Galactic average of 3.1.
The exception is HD 204827, which has an extinction curve
significantly steeper than the appropriate CCM curve.

In this paper, we take
advantage of improved IUE data along with infrared data from 2MASS and IRAS
to re-investigate the dust associated with HD 204827 and Trumpler 37. 

\section{UV Extinction Curves}

IUE spectra for all eighteen stars in Trumpler 37, previously studied
by CF, were obtained from the 
Multimission
Archive at Space Telescope (MAST). 
The archive spectra were reduced using NEWSIPS and then 
recalibrated using the method developed by Massa \& Fitzpatrick (2000). 
The signal-to-noise of the NEWSIPS IUE spectra have been improved by 10-50\% 
over those used by CF (Nichols \& Linsky 1996). 
Low dispersion LWR/LWP and SWP spectra 
were selected, 
from either aperture. Multiple spectra from one camera were averaged and then
the long and short-wavelength segments were merged at the shortest 
wavelength of the SWP. 
The wavelength coverage is $\sim$1200 -- 3200 \AA.
The sample stars and their IUE spectra are listed in Table 1. 
  
The standard pair method, in which a reddened star is compared with an 
unreddened one of the same spectral type, was used to construct each 
sightline's 
extinction curve (Massa, Savage, \& Fitzpatrick 1983). 
The unreddened comparison 
stars were selected from Cardelli, Sembach, \& Mathis (1992).
The spectral matches were made on the basis of comparing the UV spectra
of pairs of stars rather than matching their visible spectral types. 
R$_{V}$ was estimated from the JHK colors as described in
Fitzpatrick (1999).  A$_{V}$ was found using R$_{V}$ and E(B-V).
The extinction curves are normalized to A$_V$. 
Table 1 lists the sources of the UBV photometry.
Photometry in the JHK bands was availabale for all the stars in the sample 
from the 2MASS database (Cutri et al. 2003).
Table 
2 lists the spectral type, reddening, and calculated R$_{V}$ for each star 
in the sample. 
In general, our UV spectral
classifications are in good agreement with those of CF.  The resulting 
extinction curves were fit with the Fitzpatrick-Massa (FM) 
parameterization (Fitzpatrick \& Massa 1990). The fit has been limited to
the wavelength range 2700 -- 1250 \AA~(3.7-8.0 $\mu$m$^{-1}$), as it is 
not reliable longward of 2700 \AA~(E. Fitzpatrick 2002, private 
communication), and the 1250 \AA~
cut-off excludes the Ly$\alpha$ feature at 1215 \AA. 
The normalization of the FM parameters was converted from E(B-V) to A$_{V}$.
The parameters are listed in Table 3.   
In Figure 1, the extinction curve and corresponding CCM curve
for each sightline are shown.

The Cep OB2 association is a complex system with several distinct 
regions.  Previous studies have placed it at a distance of  around 800 pc 
(e.g., Garrison \& Kormendy 1976; Georgelin \& Georgelin 1976) though an 
analysis of 
Hipparcos parallaxes have placed Cep OB2 much closer, at 
615 pc (de Zeeuw et al. 1999).
Most of the stars in our sample are probably members of the Trumpler 37
association (Marschall \& van Altena 1987; de Zeeuw et al. 1999). 
The possible exceptions are HD 239724, which has been placed at
about 3 kpc (Simonson 1968) and HD 204827, placed at about 500 pc (CF). 
However, de Zeeuw et al. (1999) give a probability of 66\% that 
HD 204827 is a member of Cep OB2, and Marschall \& van Altena (1987)give a 
93\% probability 
that HD 239724
is a member of Trumpler 37.
The distance to HD 204827 is uncertain since it is a spectroscopic binary 
(Petrie \& Pearce 1961; Mason et al. 1998).

With the exception of HD 204827 and HD 239722, 
the reddenings of all the sample stars lie between 
E(B-V) of 0.4 and 0.6 mag. This reddening is primarily 
due to dust foreground to 
Trumpler 37. Using the average reddening per
kpc in the Galaxy, we would expect 0.4-0.5 mag in front of Trumpler 37
if it 
lies at a distance of 600-800 pc
(Spitzer 1973). Therefore, only the sightlines toward HD 204827 
(E(B-V) = 1.10) and HD 239722 (E(B-V)=0.93) seem to contain
significant amounts  of additional dust that
may be associated with Trumpler 37 itself.

The calculated values of R$_V$ for the sample stars tend to be smaller
than 3.1 although almost all the estimated R$_V$ values are within 2$\sigma$
of 3.1. 
Averaging the sixteen stars in the sample having 
low to moderate reddening, 
we get  E(B-V) = 0.53$\pm$0.01 mag and R$_{V}$
= 2.84$\pm$0.07. 
As can be seen in Figure 1, with the exception of HD 204827, none of the
Trumpler 37
extinction curves deviates more than 2$\sigma$ from its 
corresponding CCM curve.
The dust foreground to the Trumpler 37 appears to be 
normal diffuse interstellar dust adhering to the CCM relation. 

In an effort to separate the effects of the foreground dust and
dust local to the cloud, we partially dereddened the IUE spectra and UBVJHK
photometry for HD 204827
and HD 239722. 
A CCM-type extinction corresponding to E(B-V)= 0.55 mag and R$_{V}$
= 2.84 was removed. Extinction curves were then recalculated for these two
sightlines using these partially 
dereddened spectra, their corrected colors, and their original UV comparison 
spectra. The residual reddening toward HD 204827 from  Trumpler 37 dust 
is E(B-V) $\sim$0.55 mag and toward HD 239722 it is E(B-V) $\sim$0.4 mag. 
The new HD 239722 curve is not significantly different from the original
curve with the foreground included. They differ by only $\sim$1$\sigma$.
However, the new HD 204827 curve is significantly different. 
See Figure 2.  It is now extremely steep and 
has almost no 2175 \AA\ bump. In fact, it is indistinguishable
within the uncertainties from the average SMC bar extinction curve
(Gordon et al. 2003). This has also been plotted in Figure 2. 
The two curves lie within 1$\sigma$ of each other.  

The FM parameters for HD 204827's partially unreddened 
curve were also found. These are listed in Table 3, as are the 
average values for the SMC Bar (Gordon et al. 2003).  They were 
found with the same method Gordon et al. (2003) used to find 
FM parameters for the SMC Bar sightlines.  This required holding x$_{0}$ 
and $\gamma$ fixed at 4.60 and 1.00, respectively, while varying the 
other parameters such that $\chi^{2}$ was minimized.  Again, 
these values are within each other's uncertainties. 

\subsection{IR Emission}

Figure 3 shows 0.5$^{\circ} \times 0.5^{\circ}$ IRAS HiRes images centered on 
HD 204827 in the 25 and 60 $\mu$m bands (Aumann, Fowler \& 
Melnyk 1990). IRAS HiRes images have spatial resolution better than 1\arcmin~and
fluxes good to 20\%.
HD 204827 shows an apparent bow shock in the 25 and 60 \micron~images 
(van Buren \& McCray 1988).
Integrated fluxes in Janskys were found over a square aperture 
of 25\arcmin~on each side.  The background flux was also found and removed. 
The measured fluxes are listed in Table 4 along with the 
color temperatures estimated from the flux ratios (Ward-Thompson \& 
Robson 1991).  These are similar
to $T_{d}$ in bow shocks found elsewhere (Ward-Thompson \& Robson 1991; 
van Buren \& McCray 1988). 
We also fit 
the IRAS fluxes with a blackbody curve.  
A temperature of $\sim$75 K yielded the 
best fit for the data.  

Bow shocks are generally associated with early-type runaway stars having 
peculiar space velocities in excess of 30 km s$^{-1}$ (Blauuw 1961). 
Using the observed
radial velocity of HD 204827 (Gies 1987) and its Hipparcos proper motions,
we calculate the peculiar space velocity 
$v_{space}^{pec} = 35.1 \pm 25.2$ km/s.
Since the uncertainty in the velocity is rather large, we can only say that
the velocity HD 204827 is consistent with being a runaway star.
The standoff distance of the
HD 204827 bow shock (4.5\arcmin~or 0.9 pc at a distance of 650 pc) 
seen in Figure 3 is also consistent with this velocity 
and the standard assumptions made in Van Buren \& McCray (1988).

\subsection{IR Spectroscopy}

In 2001 and 2002 August, IR spectra of HD 204827 and HD 239722 (the other 
highly reddened cluster member) were obtained 
with the SpeX instrument at the NASA Infrared Telescope Facility (IRTF).  
SpeX is a medium resolution spectrograph which can cover a wavelength range 
from 0.8 to 5.5 $\mu$m.  For this investigation, the 1.9-4.2 $\mu$m, 
cross-dispersed mode was used. These data were reduced using version 2.0 of the 
associated SpexTool software (Cushing, Vacca, \& Rayner 2003; Rayner et al. 2003). 
To remove telluric lines, the spectrum of a ratioing G star was obtained at  
the same time, with observations bracketing those of the program star.  The 
difference in airmass between the target and the G star averaged 0.02  
and they were separated by less than one degree on the sky. The reduced target 
spectrum was then multiplied by a solar spectrum scaled to match the 
G star so that the stellar lines introduced by the division would be removed.  
The resulting spectrum then had a blackbody curve removed, corresponding 
to the effective temperature of the UV spectral classification. 
Optical depth plots were then derived following the method of Sandford et al.
(1991), by fitting and removing a linear baseline to the reduced  
spectrum between 3.23 and 3.64 $\mu$m.  For HD 204827, the feature's 
optical depth was then measured at 3.42 $\mu$m (Pendleton et al. 1994), 
yielding $\tau_{3.4}$ = 0.0139$\pm$0.0036. 

HD 204827's optical depth plot is shown in Figure 4 (top), overlayed with 
the spectrum of the Murchison meteorite (de Vries et al. 1993) and a 
zero line for easier comparison.  The 3.4 $\mu$m aliphatic C-H stretch 
is weak, but present. The middle panel of Figure 4 shows 
the optical plot with the scaled Murchison spectrum subtracted for emphasis. 
This feature arises from an organic carrier in 
the diffuse ISM. 
Pendleton et al. (1994) used a sample of sightlines, with  
A$_{V} \geq$ 3.9 mag, to show that there is a correlation between A$_{V}$ and 
the feature's optical depth $\tau_{3.4}$, with the average 
value of A$_{V}$/$\tau_{3.4}$ = 270$\pm$40 in the diffuse ISM.  
Our value of A$_{V}$/$\tau_{3.4}$ = 205$\pm$63 is in agreement with 
their results.  Thus, the sightline toward HD 204827 sets a new lower 
limit on the extinction (A$_{V}$ = 2.84$\pm0.13$) at which the feature 
has been detected.  There is no detectable 3.1 $\mu$m water ice feature 
(Pendleton et al. 1994 and references therein).
HD 239722 does not show a significant 3.4 \micron~feature, with 
$\tau_{3.4}$ = 0.010$\pm$0.006, though this sightline has a similar 
amount of extinction (A$_{V}$=2.66$\pm$0.18 mag) to HD 204827.  
Accordingly, HD 239722's A$_{V}$/$\tau_{3.4}$ ( = 266$\pm$178) is not 
significant. Its spectrum is shown in Figure 4 (bottom), again with a 
zero line for contrast.

\section{Discussion}

The residual sightline toward HD 204827 with the foreground 
dust component removed is unique 
in the Galaxy.  
About 400 hundred sightlines in the Galaxy have measured UV extinction curves
and
no other sightline in the Galaxy
shows an 
extinction curve resembling that seen in the SMC bar (Gordon \& Clayton 1998;
Valencic et al. 2003). 
This includes the sightlines near to HD 204827 in the sky which are seen 
in Figure 1.
So the HD 204827 sightline is sampling dust not
seen along the nearby sightlines toward Trumpler 37 and Cep OB2. 
For the purposes of this discussion, we shall refer to this dust 
as HD 204827 cloud dust.  Using a similar method, Whittet et al. (2003) 
find a bumpless residual dust component but with a flatter UV extinction 
toward HD 283809 in the Taurus Cloud.

HD 204827 lies in the the outer part 
of the Trumpler 37 cluster, 
away from any bright rims or areas of nebulosity, north 
northwest of IC 1396. Its projected position also lies right on the
edge of the Cepheus IRAS Bubble (Patel et al. 1998).  The IRAS 100 
$\mu$m image reveals 
that the position of HD 204827 is projected 
on the edge of a peninsula of higher optical depth 
(Abraham, Balazs, \& Kun 2000). The presence of the 
bow shock around HD 204827, indicates that the star 
may lie in or near the material
swept up in the formation of the bubble.
It formed through a combination of stellar winds and a supernova explosion
from the first generation of star formation in the region, NGC 7160 
which occurred about 7 Myr ago
(Patel et al. 1998).
The Trumpler 37 cluster formed about 5 Myr ago perhaps induced by the 
formation of the Cepheus Bubble.
Shocks such as those in supernovae ejecta will produce a dust grains size 
distribution skewed toward smaller grains. This will lead to a steeper far-UV
extinction but should also lead to a stronger 2175 \AA\ bump since this
feature is also believed to result from a population of small grains. 
(O'Donnell \& Mathis 1997).

The four sightlines in the SMC that show extinction similar to that seen 
toward the 
HD 204827 dust cloud have quite small reddenings (E(B-V)$\sim$0.2 mag). 
In addition, they are low density diffuse ISM
sightlines where the dust could have easily been subjected to 
processing by UV radiation and shocks 
(Gordon \& Clayton 1998; Gordon et al. 2003).
A group of similar 
sightlines in the Galaxy sampling very low density ISM, showed
weak bumps and steep non-CCM far-UV extinction 
(Clayton et al. 2000). But the weakness of the bumps and the 
steepness of the far-UV extinction of these sightlines do not approach that 
seen in the SMC. 
These Galactic sightlines are more similar to the those
associated with LMC2
superbubble (Misselt et al. 1999; Gordon et al. 2003). 
 
The environment of the dust in the HD 204827 cloud is quite different 
from that
seen in the SMC sightlines. 
The column density of the dust (E(B-V)=0.55 mag) 
in the HD 204827 dust
cloud is larger but the cloud also has a much higher density. The HD 204827 
dust cloud resembles a
molecular cloud more than the diffuse ISM. The cloud is very rich in
carbon molecules, showing large column
densities of C$_2$, C$_3$, CH, and CN (Oka et al. 2003; Thorburn et al. 2003).
In this respect, the HD 204827 cloud dust is quite similar to the sightline
toward HD 62542 (Cardelli \& Savage 1988). This sightline shows a severely
non-CCM sightline with a broad bump and steep far-UV 
extinction. Its dust is also rich in carbon molecules. 
The projected position of
HD 62542 lies on the edge of material swept up by a stellar wind bubble.
Three other non-CCM, weak bump, steep far-UV sightlines in the Galaxy, 
HD 283809, HD 29647, and HD 210121, are also associated 
with dense clouds (Cardelli \& Savage 1988; Larson, Whittet, \& Hough 1996; 
Cardelli \& Wallerstein 1989; Gordon et al. 2003). 
The dust in the molecular cloud 
associated with HD 210121 is likely to have been processed as it was 
propelled into the halo during a Galactic fountain or other event.  
The other two sightlines toward HD 29647 and HD 283809 seem 
to be sampling dust in quiescent dense clouds. These two sightlines
show a strong 3.1 \micron~ice feature and a weak 3.4 \micron~feature similar
to the one seen in the HD 204827 cloud dust (Goebel 1983; Smith, Sellgren, 
\& Brooke 1993). 
The steep far-UV extinction in these clouds 
helps shield the molecules in these clouds
from dissociating UV radiation leading to larger column densities
than might be found in clouds with less steep far-UV extinction (Mathis 1990). 

So the conditions for producing SMC-type extinction exist in our own galaxy.
Those conditions are not necessarily associated with low reddening, 
low density, diffuse ISM environments. Also, metallicity differences between 
the Galaxy and the SMC may not be a determining factor. This supports
results that find SMC-type extinction in starburst galaxies
having a wide range of metallicities (Calzetti, Kinney, \& Storchi-Bergmann 
1994; Gordon et al.1997).

\newpage

\begin{center}
Table~1 \\[0.1cm]
IUE Spectra and Photometry Sources of Program Stars \\ [0.1cm] 
\begin{tabular}{rcccc} \tableline \tableline
HD/BD & SWP & LWP/LWR &  UBV Source \\ \tableline 
204827 & 11131, 14530 & 09761, 11104 &  1 \\
205794 & 23119 & 03451 & 2  \\
205948 & 23122 & 03453 & 2  \\
206267 & 26011 & 06057, 17971 & 2  \\
239683 & 23130 & 03460 & 2 \\
239689 & 23138 & 03466 & 2 \\
239693 & 23131 & 03463 & 2 \\
239710 & 17470 & 13757 & 2 \\
239722 & 23125 & 03456 & 3 \\
239724 & 23118 & 03450 & 2 \\
239725 & 23136 & 03464 & 2 \\
239729 & 13451 & 10114 & 2 \\
239738 & 23123 & 03454 & 3 \\
239742 & 23129 & 03459 & 4 \\
239745 & 23128 & 03458 & 2 \\
239748 & 23137 & 03465 & 4 \\
+57 2395B & 23132 & 03461 & 3 \\ 
+58 2292 & 23139 & 03467 & 3 \\ \tableline 
\end{tabular}
\end{center}
\vspace{5 mm}
NOTE. -- (1) Hiltner 1956, (2) Nicolet 1978, (3) Garrison \& Kormendy 1976, (4) Simonson 1968.  All JHK photometry is from 2MASS.  

\newpage

\begin{center}
Table~2 \\[0.1cm]
Properties of Program Stars \\ [0.1cm]
\begin{tabular}{rllccr}  \tableline \tableline
HD/BD & Sp Type & UV Sp Type & E(B-V) & R$_{V}$ \\ \tableline 
204827 & B0 V & B0 V & 1.10$\pm$0.05 & 2.58$\pm$0.12 \\
205794 & B5 V & B0.5 V & 0.62$\pm$0.05 & 3.09$\pm$0.26 \\ 
205948 & B2 V & B1 V & 0.50$\pm$0.04 & 2.90$\pm$0.27 \\
206267 & O6 V & O7 V & 0.52$\pm$0.04 & 2.82$\pm$0.22 \\
239683 & B5 V & B2 IV & 0.54$\pm$0.04 & 2.76$\pm$0.22 \\
239689 & B5 V & B1 V & 0.45$\pm$0.04 & 2.70$\pm$0.29 \\
239693 & B5 V & B4 IV & 0.41$\pm$0.04 & 2.37$\pm$0.27 \\
239710 & B3 V & B1 V & 0.62$\pm$0.07 & 3.02$\pm$0.32 \\
239722 & B5 V & B1 V & 0.93$\pm$0.05 & 2.86$\pm$0.17 \\
239724 & B1 III & B1.5 III & 0.62$\pm$0.04 & 3.18$\pm$0.24 \\
239725 & B5 V & B1 V & 0.52$\pm$0.04 & 3.14$\pm$0.28 \\
239729 & B0 V & O9 V & 0.66$\pm$0.04 & 3.19$\pm$0.19 \\
239738 & B5 V & B2 V & 0.51$\pm$0.05 & 2.90$\pm$0.32 \\
239742 & B5 V & B4 IV & 0.38$\pm$0.04 & 2.36$\pm$0.31 \\
239745 & B5 V & B1.5 V & 0.54$\pm$0.07 & 2.66$\pm$0.34 \\
239748 & B5 V & B1 V & 0.43$\pm$0.04 & 2.93$\pm$0.31 \\
+57 2395B & B5 V & B2 V & 0.64$\pm$0.04 & 2.44$\pm$0.19 \\
+58 2292 & B5 V & B2 V & 0.57$\pm$0.03 & 3.00$\pm$0.26 \\ \tableline 
\end{tabular}
\end{center}

\newpage

\begin{center}
Table~3 \\[0.1cm]
FM Parameters of Program Stars \\ [0.1cm]
\begin{tabular}{lcccccc} \tableline \tableline
HD/BD & c$_{1}$/R$_{V}$ & c$_{2}$/R$_{V}$ & c$_{3}$/R$_{V}$ & c$_{4}$/R$_{V}$ & x$_{0}$ & $\gamma$ \\ \tableline 

204827 & 1.08$\pm$0.38 & 0.38$\pm$0.03 & 0.66$\pm$0.13 & 0.36$\pm$0.06 & 4.66$\pm$0.02 & 0.91$\pm$0.03 \\
205794 & 1.27$\pm$0.40 & 0.17$\pm$0.03 & 1.05$\pm$0.19 & 0.14$\pm$0.04 & 4.57$\pm$0.01 & 0.88$\pm$0.02 \\
205948 & 1.36$\pm$0.43 & 0.16$\pm$0.03 & 1.17$\pm$0.25 & 0.18$\pm$0.04 & 4.59$\pm$0.01 & 0.91$\pm$0.03 \\
206267 & 1.17$\pm$0.45 & 0.27$\pm$0.04 & 1.02$\pm$0.20 & 0.22$\pm$0.05 & 4.59$\pm$0.01 & 0.91$\pm$0.03 \\
239683 & 1.21$\pm$0.43 & 0.23$\pm$0.04 & 1.59$\pm$0.39 & 0.24$\pm$0.07 & 4.59$\pm$0.03 & 1.18$\pm$0.04 \\
239689 & 0.84$\pm$0.27 & 0.28$\pm$0.04 & 1.29$\pm$0.18 & 0.18$\pm$0.03 & 4.57$\pm$0.005 &  0.96$\pm$0.02 \\
239693 & 1.16$\pm$0.43 & 0.23$\pm$0.05 & 1.13$\pm$0.26 & 0.20$\pm$0.06 & 4.57$\pm$0.95 & 0.90$\pm$0.03 \\
239710 & 1.56$\pm$0.49 & 0.15$\pm$0.04 & 0.71$\pm$0.18 & 0.20$\pm$0.06 & 4.60$\pm$0.02 & 0.82$\pm$0.03 \\
239722 & 0.88$\pm$0.34 & 0.32$\pm$0.04 & 1.28$\pm$0.21 & 0.22$\pm$0.04 & 4.59$\pm$0.01 & 1.04$\pm$0.03 \\
239724 & 1.09$\pm$0.35 & 0.21$\pm$0.03 & 1.07$\pm$0.19 & 0.14$\pm$0.03 & 4.60$\pm$0.01 & 0.94$\pm$0.03 \\
239725 & 1.16$\pm$0.42 & 0.22$\pm$0.03 & 0.87$\pm$0.16 & 0.21$\pm$0.04 & 4.56$\pm$0.01 & 0.91$\pm$0.03 \\
239729 & 1.11$\pm$0.25 & 0.22$\pm$0.02 & 1.01$\pm$0.16 & 0.25$\pm$0.04 & 4.61$\pm$0.01 & 1.08$\pm$0.03 \\
239738 & 1.00$\pm$0.01 & 0.24$\pm$0.04 & 1.17$\pm$0.28 & 0.20$\pm$0.06 & 4.55$\pm$0.01 & 1.03$\pm$0.03 \\
239742 & 0.57$\pm$0.33 & 0.38$\pm$0.11 & 1.84$\pm$0.58 & 0.15$\pm$0.07 & 4.58$\pm$0.02 & 1.03$\pm$0.04 \\
239745 & 1.12$\pm$0.28 & 0.22$\pm$0.04 & 1.36$\pm$0.31 & 0.19$\pm$0.05 & 4.54$\pm$0.004 & 0.93$\pm$0.02 \\
239748 & 1.21$\pm$0.41 & 0.20$\pm$0.04 & 1.08$\pm$0.25 & 0.25$\pm$0.06 & 4.57$\pm$0.01 & 0.88$\pm$0.03 \\
+57 2395B & 1.28$\pm$0.41 & 0.35$\pm$0.05 & 1.29$\pm$0.24 & 0.31$\pm$0.06 & 4.57$\pm$0.01 & 0.97$\pm$0.03 \\ 
+58 2292 & 0.94$\pm$0.29 & 0.25$\pm$0.04 & 1.08$\pm$0.18 & 0.14$\pm$0.03 & 4.58$\pm$0.01 & 0.92$\pm$0.02 \\ \tableline 
204827 &  -1.85$\pm$0.54 & 0.82$\pm$0.15 & 0.24$\pm$0.09 & 0.11$\pm$0.04 & 4.60$\pm$0.00 & 1.00$\pm$0.00 \\ 
(partially dereddened) & & & & & & \\ 
SMC Bar average & -1.81$\pm$0.16 & 0.83$\pm$0.15 & 0.14$\pm$0.05 & 0.17$\pm$0.02& 4.60$\pm$0.00 & 1.00$\pm$0.00 \\

\tableline 
\end{tabular}
\end{center}

\newpage

\begin{center}
Table~4 \\[0.1cm]
IRAS Fluxes and Color Temperatures \\ [0.1cm]
\begin{tabular}{ccccccccc} \hline \hline
& F(12$\mu$m) & F(25$\mu$m) & F(60$\mu$m) & F(100$\mu$m) & T$_{12/25}$ & T$_{25/60}$ & T$_{60/100}$\\ 
& (Jy) & (Jy) & (Jy) & (Jy) & (K) & (K) & (K) \\ \tableline 
HD 204827 & .14 & 27.84 & 170.27 & 146.03 & 70 & 55 & 38\\ 
 & $\pm$0.08 & $\pm$1.11 & $\pm$4.43 & $\pm$14.84 & $\pm$7 & $\pm$1 & $\pm$3 \\ \tableline 
\end{tabular}
\end{center}

\newpage

\begin{figure}[th]
\begin{center}
\plotone{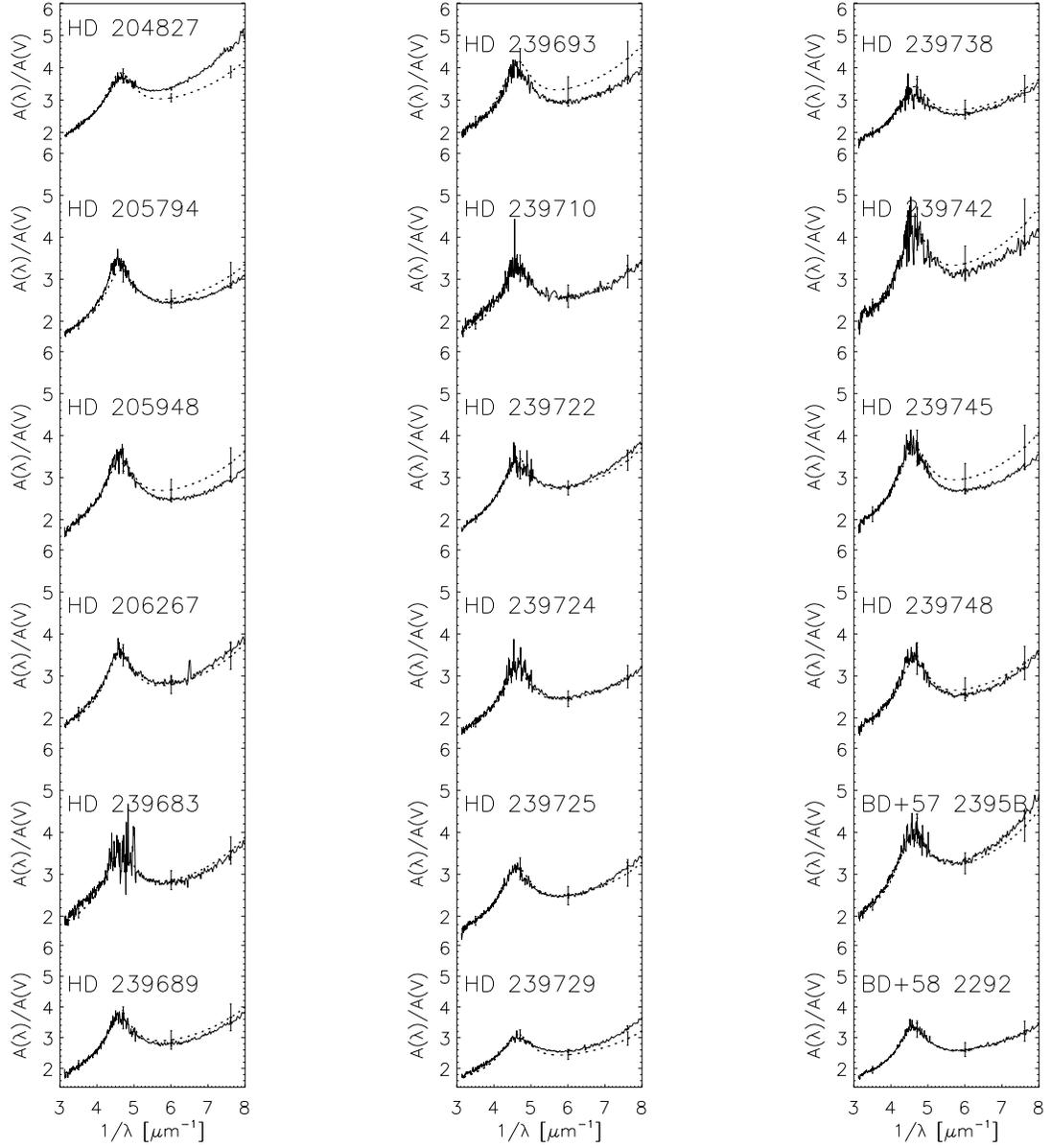}
\end{center}
\caption{The extinction curves of the Trumpler 37 sightlines overlayed with 
CCM curves appropriate to R$_{V}$ along each sightline.  The error bars indicate a 1$\sigma$
uncertainty.}
\end{figure}

\begin{figure}[th]
\begin{center}
\plotone{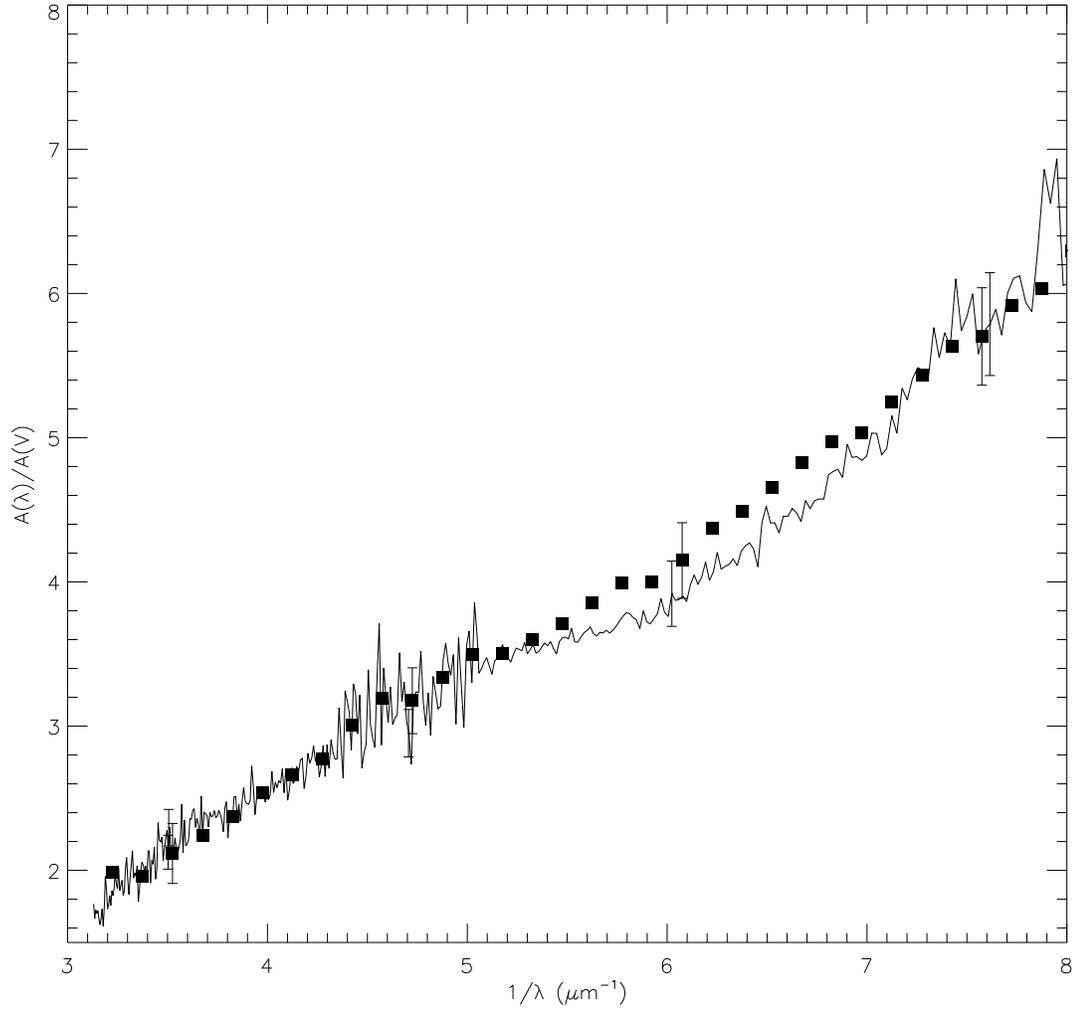}
\end{center}
\caption{The partially dereddened extinction curve for HD 204827 (solid 
line). The extinction curve to the the SMC bar (squares) is from Gordon et al. 
(2003) is also shown. The error bars indicate a 1 $\sigma$ uncertainty.}  
\end{figure}

\begin{figure}[th]
\begin{center}
\plottwo{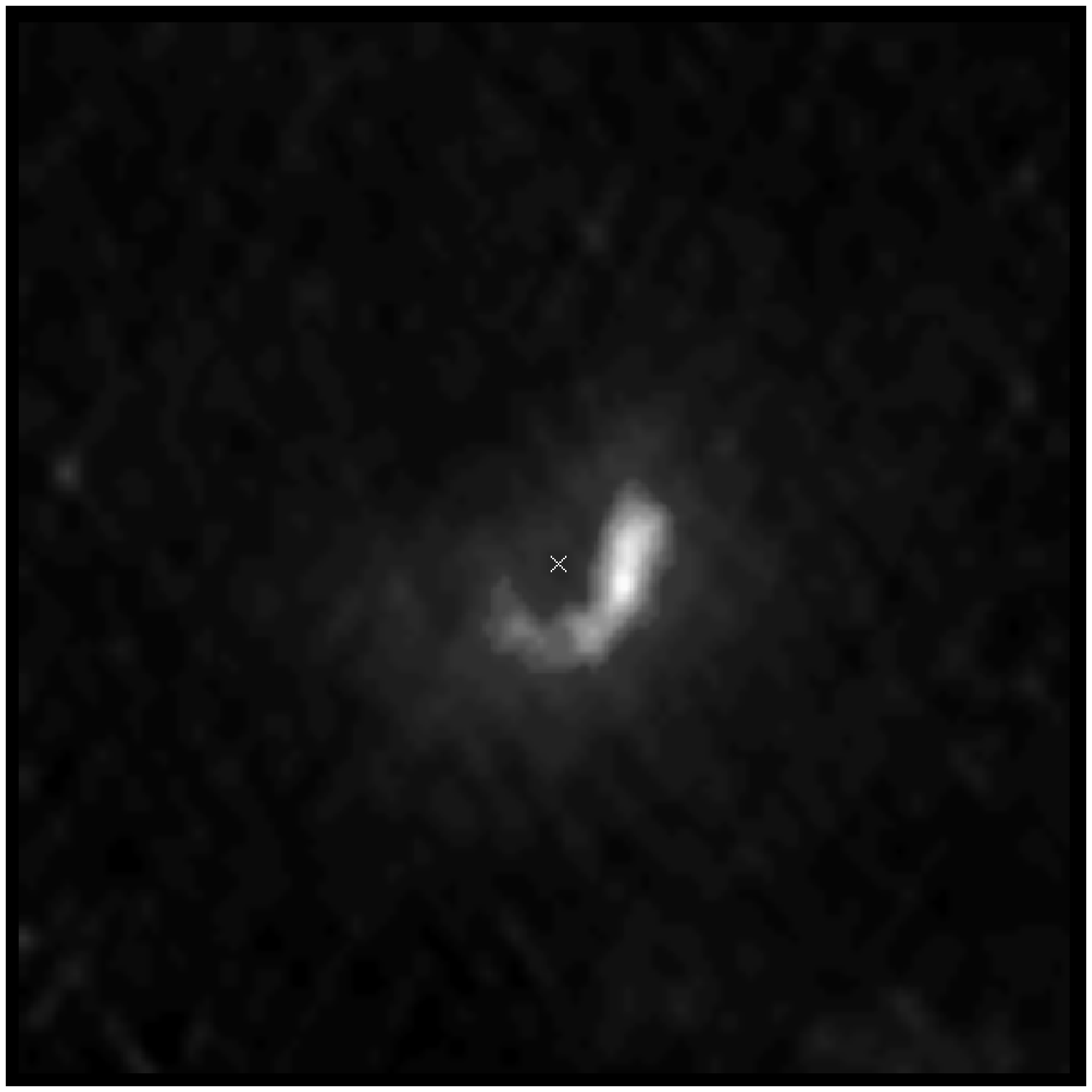}{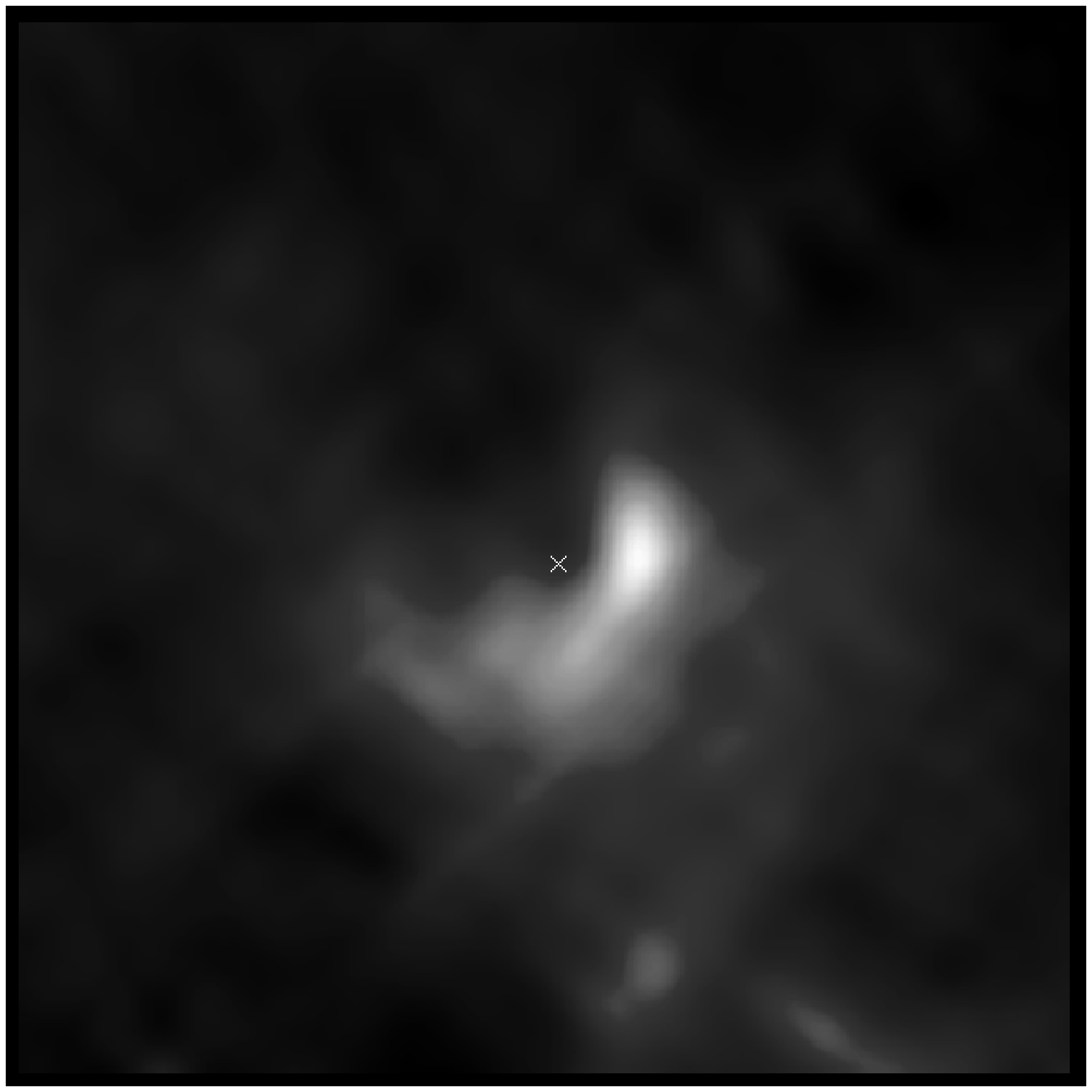}
\end{center}
\caption{0.5 $^{\circ} \times 0.5 ^{\circ}$ IRAS HiRes images of HD 204827 at 
60 (left) and 100 $\mu$m (right), respectively.  The x indicates the star's location.}
\end{figure}

\begin{figure}[th]
\begin{center}
\plotone{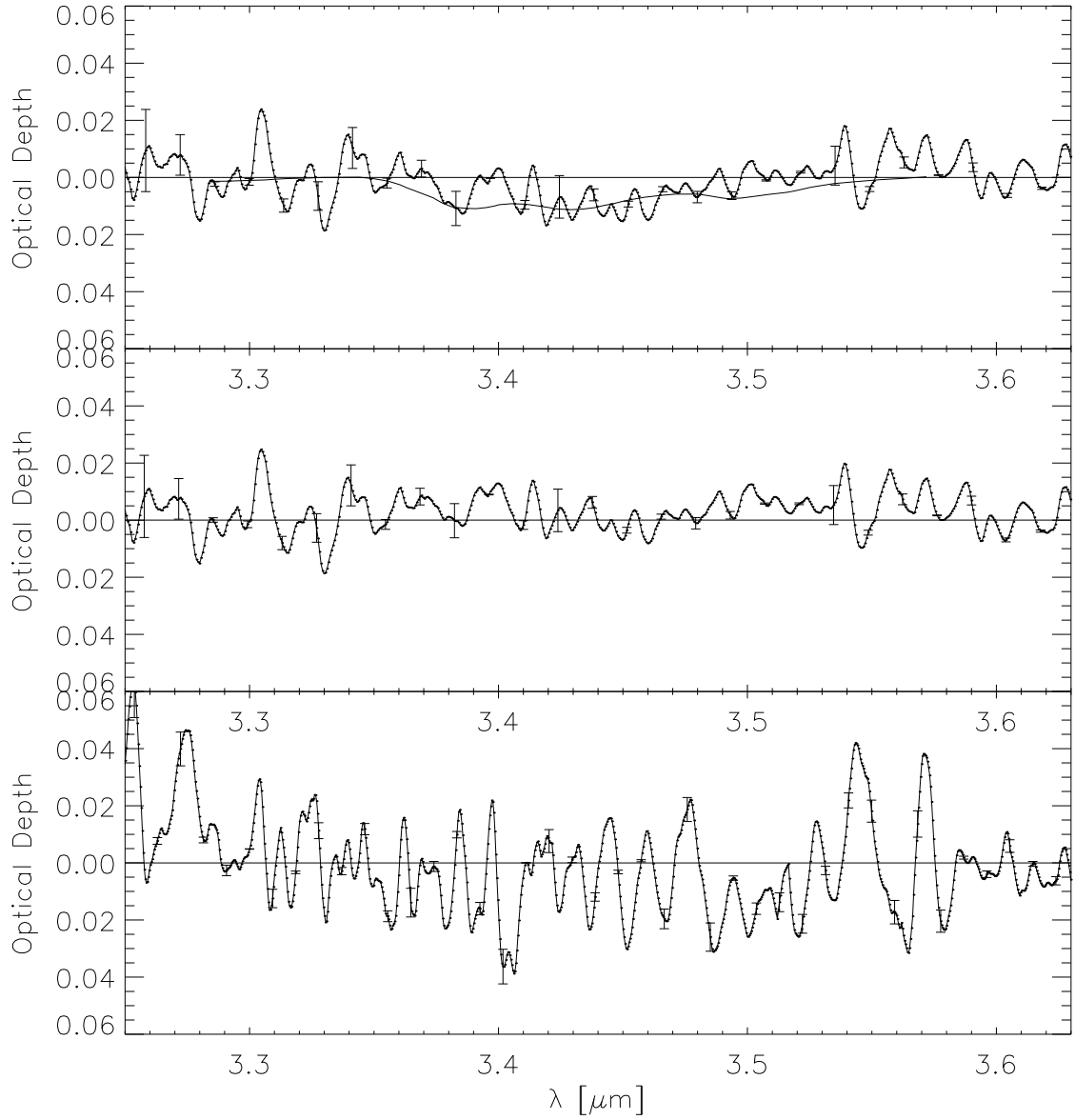}
\end{center}
\caption{
Optical depth plots of HD 204827 and HD 239722 for comparison.
Both were obtained with SpeX.  Top: HD 204827 with the C-H aliphatic stretch
at 3.4 $\mu$m overlayed (de Vries et al. 1993).  Middle: HD 204827 with the
3.4 $\mu$m feature removed for emphasis.  Bottom: HD 239722; no feature is
visible. The error bars represent 1 $\sigma$ uncertainty.}
\end{figure}

\end{document}